# Possibility to construct a clock on non 0-0 transition*

Chao Li[1,2], Fuyu Sun[1,**], Jie Liu[1], Xiaofeng Li[1], Jie Ma[1], Guangkun Guo[3], Dong Hou[3], Shougang Zhang[1]

[1] Key Laboratory of Time and Frequency Primary Standards, National Time Service Center, Chinese Academy of Sciences, Xi'an 710600, China

[2] University of Chinese Academy of Sciences, Beijing 100049, China

[3] Time & Frequency Research Center, School of Automation Engineering, University of Electronic Science and Technology of China, Chengdu 611731, China

*Supported by the National Natural Science Foundation of China under Grant Nos. 61871084 and 61601084.

**To whom correspondence should be addressed. Email: sunfuyu@ntsc.ac.cn

**Abstract** In this letter, we discuss the possibility of non 0-0 atomic lines as the reference transitions to design a new atomic clock. The proposed clock operates in high C-field regime offering an interesting alternative to conventional clock based on 0-0 transition as the former can also eliminate the influence of the first-order Zeeman effect. Using the Breit-Rabi formula, we theoretically calculate the magnetically insensitive frequency and the corresponding C-field value for different alkali atom species. We also provide an estimate of the uncertainty of frequency shift to change of C-field. We show that a high C-field clock based on Cs: $|3,-1\rangle \leftrightarrow |4,0\rangle$, $|3,0\rangle \leftrightarrow |4,-1\rangle$, and $|3,-1\rangle \leftrightarrow |4,-1\rangle$, can achieve a bias of $10^{-14}$ with a C-field of $10^{-6}$ uncertainty. The free first-order Zeeman effect C-field for $|3,-1\rangle \leftrightarrow |4,0\rangle$ or $|3,0\rangle \leftrightarrow |4,-1\rangle$ transition is about one half that for $|3,-1\rangle \leftrightarrow |4,-1\rangle$ transition.

Time and frequency are the most precisely measured physical quantities,[1,2] which, apart from practical applications in timekeeping, synchronization, and information infrastructures, play critical roles in tests of fundamental physics.[3,4] Based on this, atomic clocks that traditionally designed to probe the hyperfine transition in the electronic ground state of the alkali-metal atom, have been researched extensively in the last few decades. One of the central requirements for developing a high-performance atomic clock is to make the clock frequency insensitive to the fluctuation in static magnetic field. With this consideration, many theoretical and experimental studies have been devoted to magnetically fluctuation-immune 0-0 transition as usually employed in traditional atomic clocks.[5]

Motivated by the early work of De Marchi and co-workers,[6-8] we here investigate in detail the possibility of exciting the clock transition between non 0-0 states with which the first-order Zeeman shift is exactly cancelled. Before discussing this, we firstly recall the traditional case. The well-known Breit-Rabi formula tell us the relevant energy levels of alkali atoms placed inside a static magnetic field $B$, here $B$, i.e., C-field for simplicity, defining a spin-quantization axis for atoms. Thus the frequency of π transition ($\Delta F = 1$, $\Delta m_F = 0$) reads $v_\pi = v_{00} \left(1 + \frac{4m_F x}{2I+1} + x^2\right)^{1/2}$, here $x = \frac{(g_J - g_I)\mu_B}{h v_{00}} B$, $h$ the Plank constant, $v_{00}$ the unperturbed ground-state hyperfine transition frequency, $I$ the nuclear spin, $g_J$ and $g_I$ are the electronic and nuclear g-factors under investigation, respectively, $\mu_B$ the Bohr magneton. Obviously, the Zeeman effect can be significantly suppressed by choosing the π transition among $m_F = 0$ states and operating the clock in weak C-field regime ($x^2 \ll 1$). Taking Cs atom as a calculating example, the $|3,0\rangle \leftrightarrow |4,0\rangle$ transition frequency is first-order magnetic field independent, whereas the frequency of the six others is a linear function of applied C-field. In weak C-field regime, the error relating to the Zeeman frequency shift is about $\delta v/v_{00} \approx 3.46 \times 10^{-10}$ $\delta B/B$ implying that a change of $\delta B \sim 1$ nT will cause a frequency bias of $\delta v/v_{00} \sim 5.77 \times 10^{-14}$ while $B$ is stabilized at ~6 μT. This is less challenging, as a consequence, most often the stabilization of atomic frequency is achieved via ground state 0-0 transition.

Aiming to investigate the effect of a high C-field on hyperfine frequency, we here consider from a general view of the choose of magnetic field insensitive (free first-order Zeeman effect) transition and rewrite the expression of ground state π transition as follows: $v_\pi = v_{00}\left[1 - \left(\frac{2m_F}{2I+1}\right)^2 + \left(x + \frac{2m_F}{2I+1}\right)^2\right]^{1/2}$, then we find that the first-order Zeeman effect can be exactly cancelled when the following condition is satisfied for the case of $m_F < 0$:

$$B = -\frac{2m_F}{2I+1}\frac{hv_{00}}{(g_J - g_I)\mu_B}$$

This relationship allows us to quickly estimate the needed C-field that makes the atomic transition insensitive to field fluctuation, thereby greatly reducing the frequency bias. Under this condition, the clock frequency reads $v_{00}\left[1 - \left(\frac{2m_F}{2I+1}\right)^2\right]^{1/2}$ and the residual second-order Zeeman frequency shift can be estimated as

$$\delta v = \frac{[(g_J - g_I)\mu_B]^2}{2h^2 v_{00}\sqrt{1 - \left(\frac{2m_F}{2I+1}\right)^2}}(\delta B)^2$$

Following the similar strategy we calculate the magnetically insensitive atomic frequency of non 0-0 π transition and the corresponding C-field value for various alkali metal atoms as presented in Table I, where $\delta B$ is expressed in teslas. These investigations allows us to identify the special C-field which minimizes the frequency variation, enabling a new clock with high C-field. The key to realizing such a clock is to ensure that the C-field value is exactly set to the Breit-Rabi minimum.

**TABLE I.** A list of examples of selectable, potential free first-order Zeeman effect π transitions ($\Delta F = 1$, $\Delta m_F = 0$, $m_F < 0$) for high C-field operation.

| Data for | $^{23}$Na | $^{39}$K | $^{87}$Rb | $^{85}$Rb | | $^{133}$Cs | | |
|---|---|---|---|---|---|---|---|---|
| $v_{00}$(GHz) | ~1.7716 | ~0.4617 | ~6.8347 | ~3.0357 | | ~9.1926 | | |
| $I$ | 3/2 | 3/2 | 3/2 | 5/2 | | 7/2 | | |
| $m_F$ | -1 | -1 | -1 | -1 | -2 | -1 | -2 | -3 |
| $B$(G) | ~316 | ~82 | ~1219 | ~361 | ~722 | ~820 | ~1640 | ~2459 |
| $v_\pi$(GHz) | $\frac{\sqrt{3}}{2}v_{00}$ | $\frac{\sqrt{3}}{2}v_{00}$ | $\frac{\sqrt{3}}{2}v_{00}$ | $\frac{2\sqrt{2}}{3}v_{00}$ | $\frac{\sqrt{5}}{3}v_{00}$ | $\frac{\sqrt{15}}{4}v_{00}$ | $\frac{\sqrt{3}}{2}v_{00}$ | $\frac{\sqrt{7}}{4}v_{00}$ |
| $\frac{\delta v}{v_\pi}$ | ~166.95 × $(\delta B)^2$ | ~2456.66 × $(\delta B)^2$ | ~11.22 × $(\delta B)^2$ | ~47.95 × $(\delta B)^2$ | ~76.73 × $(\delta B)^2$ | ~4.96 × $(\delta B)^2$ | ~6.20 × $(\delta B)^2$ | ~10.63 × $(\delta B)^2$ |

Table I indicates that non 0-0 transition could also be chosen as reference transition by applying a specially-designed C- field. Around the optimum C-field, the promising transition has a second-order Zeeman shift proportional to $(\delta B)^2$. The transition in high field regime possess several advantages over 0-0 transition in weak field regime.[8] First, the influence of neighboring transitions is decreased to a negligible level. Second, the requirement for magnetic shielding is significantly reduced. Finally, some applications involving high field precision spectroscopy could operate at these specific values, thus minimizing magnetic fluctuation.

From the analysis above, one can calculate the magnetically insensitive C-field to be ~ 820 G[6-8] for Cs: $|3,-1\rangle \leftrightarrow |4,-1\rangle$ at where the Zeeman dependent uncertainty drops below $5\times10^{-14}$ in a $10^{-6}$ C-field stabilization level; the transition span is enlarged by several orders of magnitude (from tens of kHz to hundreds of MHz); two orders of magnitude of magnetic fluctuation larger than 0-0 transition is allowable, to achieve the same accuracy of the order of $10^{-14}$. In addition to the three π transitions listed in Table I, there exist six σ transitions fulfilling the condition of first-order cancellation of Zeeman shift. All the selectable hyperfine transitions supporting magnetically insensitive operation and their frequency dependences for different applied C-field are shown in Fig. 1, including 0-0 transition.

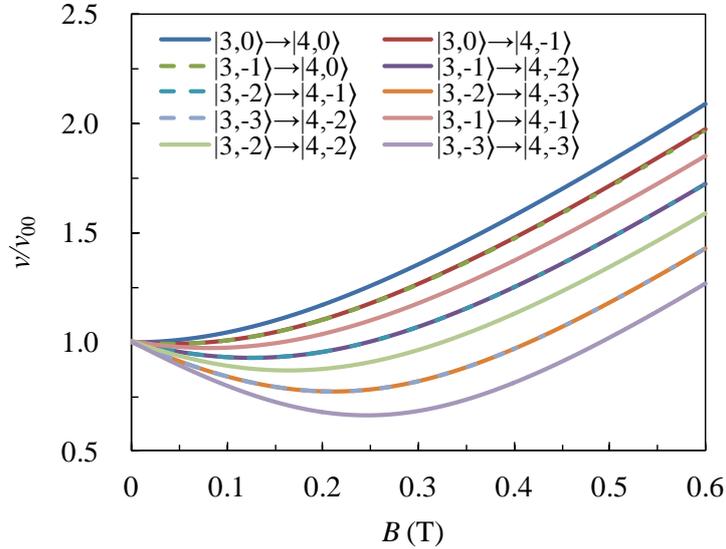

Fig. 1. Atomic transition frequency as a function of C-field for the Cs hyperfine ground states. There are total nine allowed field-independent transitions satisfying the condition of zero derivative of the frequency with respect to C-field. The 0-0 transition is shown for comparison.

It is worth noting that obtaining a ultrastable high C-field become more challenging than the weak field case.[9,10] From a practical point of view, the more applicable way to excite non 0-0 clock transition is to choose those transitions involving lower Zeeman states. From Fig. 1, we point out that the ground state $\sigma_-: |3, 0\rangle \leftrightarrow |4, -1\rangle$ and $\sigma_+: |3, -1\rangle \leftrightarrow |4, 0\rangle$ transitions are worthy of our attention as their magnetically insensitive points (~ 415.8 G and 417.2 G) appear when the C-field is about only half that of $\pi: |3, -1\rangle \leftrightarrow |4, -1\rangle$ transition. Notice that the magnetically insensitive points between the two $\sigma$ transitions above are very close to overlapping. The corresponding C-field difference is less than 2 G.

In order to study more details about the non 0-0 states, free first-order Zeeman transitions and to investigate how the magnetic fluctuation affects the frequency bias. We numerically simulate the uncertainty in atomic frequency induced by the change in C-field setting a common target accuracy of ~ $5 \times 10^{-14}$ as shown in Fig. 2. The inset of Fig. 2 shows the dependence of frequency uncertainty on C-field fluctuation.

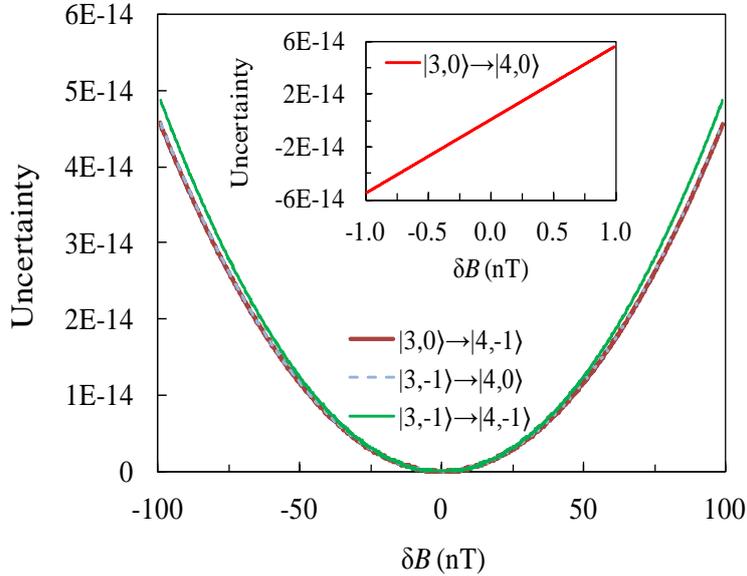

Fig. 2. $\delta\upsilon/\upsilon_\pi$ and $\delta\upsilon/\upsilon_\sigma$ vs $\delta B$ around optimum C-field eliminating the first-order magnetically induced frequency shift. The optimum C-field strengths for $|3, 0\rangle \leftrightarrow |4, -1\rangle$, $|3, -1\rangle \leftrightarrow |4, 0\rangle$, and $|3, -1\rangle \leftrightarrow |4, -1\rangle$ transitions are ~ 415.8 G, 417.2 G, and 819.8 G, respectively. Inset: $\delta\upsilon/\upsilon_{00}$ vs $\delta B$ when $B = 6\,\mu T$.

Setting a C-field dependent uncertainty given in Fig. 2, one could find that the clock transition using $|3,0\rangle \leftrightarrow |4,-1\rangle$ or $|3,-1\rangle \leftrightarrow |4,0\rangle$ have an advantage of reduced optimum C-field over De Marchi transition[8] using $|3,-1\rangle \leftrightarrow |4,-1\rangle$ while maintaining the almost identical frequency uncertainty. More importantly, exciting the three high field non 0-0 transitions, compared to conventional weak field 0-0 transition, has led to a significant increases in the tolerance of absolute magnetic fluctuation by two orders of magnitude (2 nT → 200 nT). This implies a considerable reduction in magnetic shielding performance. Besides, we emphasize that the requirement for relative magnetic fluctuation in high C-field case is extremely stringent ($10^{-4} \to 10^{-6}$), which is the major limitation of this operation in the current implement. Recently, a stability better than 0.1 ppm/h under a 10 A DC current is reported to be achievable,[10,11] allowing the preliminary new clock test in short term.

The well-known Breit-Rabi formula provide us the value of atomic transition frequency with a enough precision in weak and intermediate magnetic fields, meeting the requirements of most current scientific and engineering applications. This is not enough for higher accuracy in high field condition. Some additional frequency shift terms should be carefully treated beyond the traditional Zeeman relating problem discussed above.[12-14] As firstly pointed out by Itano, at least three corrections to the Breit-Rabi formula should be evaluated[15]—the hyperfine-assisted Zeeman shift, the dipole diamagnetic shift, and the quadrupole diamagnetic shift. Itano theoretically estimated these deviations from the Breit-Rabi prediction and showed that about a few parts in $10^{12}$ total shift is involved for π: $|3,-1\rangle \leftrightarrow |4,-1\rangle$ transition.[15] When choosing σ transition as a reference line, however, we should also carefully consider the other contribution, e.g., Millman effect relating frequency shift.[5] According to the discussion above, we plan to construct a beam-type[16-18] high C-field clock based on the non 0-0 transition. As first steps toward this new clock, we are performing the preliminary design of Ramsey microwave cavity[19] and ultrastable DC source.

In summary, much of the focus on the clock transition has been for the ground states 0-0 transition in weak field regime. Here we present a study of non 0-0 transition operating in high field regime, which are able to exactly eliminate the first-order Zeeman effect. We investigate the frequency bias of non 0-0 transition under first-order Zeeman cancellation in terms of the C-field change for various alkali atom species using Breit-Rabi formula and discuss their advantages over traditional 0-0 transition. Our results indeed suggest a possibility of effectively stabilizing an atomic frequency with the use of $|F, 0\rangle \leftrightarrow |F \pm 1, -1\rangle$ or $|F, -1\rangle \leftrightarrow |F + 1, -1\rangle$ transition. Any improvement in C-field will directly improve accuracy. The combination of a high homogeneity C-field scheme and a ultrastable DC source makes the discussed transition a promising candidate for high-performance atomic clock and other applications involving magnetic-immune precision spectroscopy.